\documentclass[doublecol]{epl2-per}

\title{Vortex density spectrum of quantum turbulence}
\shorttitle{Vortex spectrum of quantum turbulence} 

\author{P.-E. Roche\inst{1} \and P. Diribarne\inst{1} \and T. Didelot\inst{1} \and O. Fran\c{c}ais\inst{2,3} \and L. Rousseau\inst{2} \and H. Willaime\inst{4}}
\shortauthor{P.-E. Roche \etal}

\institute{                    
  \inst{1} Institut N\'eel, Centre National de la Recherche Scientifique - \\25 avenue des Martyrs, BP166, 38042 Grenoble Cedex 9, France\\
  \inst{2} Groupe ESIEE, Chambre de Commerce et d'Industrie de Paris - \\Cit\'e Descartes BP99 - 2 boulevard Blaise Pascal, 93162 Noisy le Grand Cedex, France\\
    \inst{3} Conservatoire National des Arts et M\'etiers -  Laboratoire de Physique, 2 rue Cont\'e, 75003 Paris, France\\
    \inst{4} Ecole Sup\'erieure de Physique et de Chimie Industrielles de la Ville de Paris -\\ 10 rue Vauquelin, 75231 Paris Cedex 05, France 
}

\abstract{
The fluctuations of the vortex density in a turbulent quantum fluid are deduced from local second-sound attenuation measurements. These measurements are performed with a micromachined open-cavity resonator inserted across a flow of turbulent He-II near 1.6 K.  The power spectrum of the measured vortex line density is compatible with a (-5/3) power law. The physical interpretation, still open, is discussed.}

\begin{document}

\PSLogo{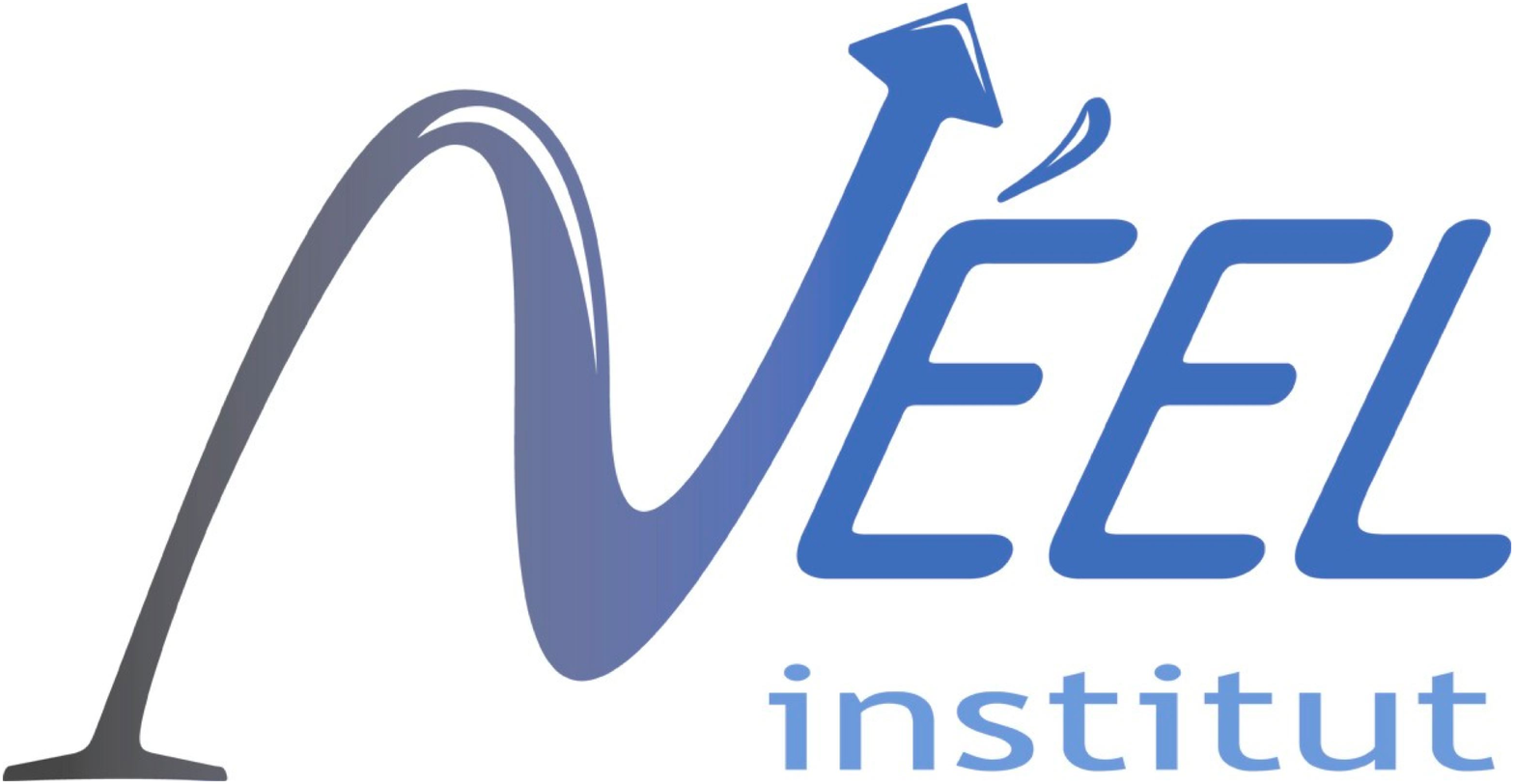}

\maketitle

\section{Motivation}

Our understanding of the turbulent fluctuations in quantum fluid flows relies on a single direct measurement\cite{Maurer1998}. This lack of local experimental data\cite{VinenJLTP2001} contrasts with the situation for counter-flow turbulence, which has been explored by a dozen of fluctuations studies (\cite{Tough1982, DonnellyLivreVortices,Nemirovskii1995}). It also contrasts with the detailed theoretical picture of the Kolmogorov-like cascade which emerged over the last decades \cite{Vinen2002} and with the diversity of numerical simulations, such as velocity spectrum studies \cite{Nore1997,KivotidesSpectra2002,KobayashiPRL2005}.

In the isolated experiment mentioned above\cite{Maurer1998}, a miniature total-head tube was configured to operate as an anemometer in a He-II co-flow (or ``bulk flow''), at $2.08$ and $1.4\,K$. In a two-fluid model picture, this probe senses a combinaison of the velocities of the superfluid and normal components. Its spatial and time resolution (typically $2\,mm$ and $1\,kHz$) enabled to resolve one decade of power law scaling below the injection scale. This scaling was found to overlap with the scaling in the inertial range of classical flows. This strong result suggests that the largest scales structures in quantum turbulence undergo a Kolmogorov-type cascade.

Second sound probes have a several decades history as an efficient tool to measure quantum vortices line density (VLD), in particular in turbulent co-flows (for instance \cite{Holmes1992, Smith1993,stalp2002} and references within) and counter-flows (for instance  \cite{Tough1982, DonnellyLivreVortices,Nemirovskii1995,Ostermeier1980,skrbek2003} and references within). Unfortunately, the size of these sensors and their sidewall positionning made impossible space and time resolved measurements of flow fluctuations. The aim of this paper is to report such a local fluctuations measurement from a micromachined miniature second sound resonator. This probe therefore completes the inertial range characterization with a fully superfluid observable, independent of the earlier velocity measurements. 

\section{The flow set-up}

\begin{figure}
\onefigure[width=6.5cm]{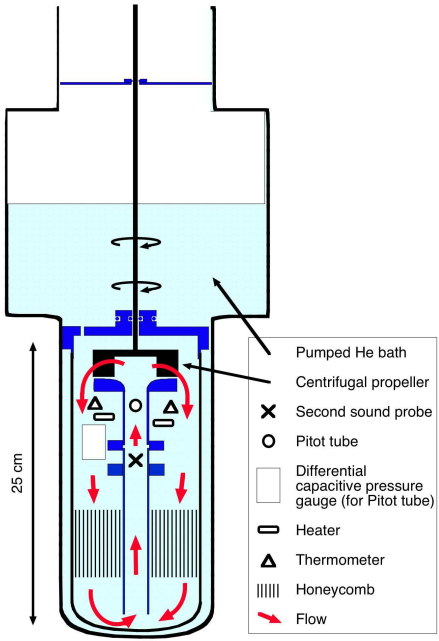} 
\caption{
Scheme of the flow loop with probes.
\label{fig:flow}}
\end{figure}

Our flow is a He-II  loop confined in a nearly cylindrical cryostat and continuously powered by a centrifugal pump (see figure~\ref{fig:flow}). Turbulence is probed in a $\Phi = 23\,mm$-diameter, $250\,mm$-long brass pipe, located upstream from the pump . Downstream the pump, the fluid returns to the pipe inlet flowing along the outside of it. On this return path, a $30\,mm$-long $3\,mm$-cell honeycomb breaks the spin motion generated by the pump. As a test, another $20\,mm$-long honeycomb has once been inserted in the pipe inlet without noticeable changes. A Pitot tube is located $130\,mm$ before the pipe outlet. It provides a measurement of the mean velocity by mean of an in-situ capacitive differential pressure gauge. 
The usefull range of velocity is $V=0.3-1.3\,m/s$. At lower velocities flow instabilities are detected and at higher velocities, typically at $1.5\,m/s$, a cavitation threshold is encountered. From in-situ measurements with a semiconducting miniature pressure sensors\cite{HaruyamaAdv1998}, we estimate a typical $35 \%$ velocity turbulence ratio in the pipe.
With velocity $V$ and pipe diameter $\Phi$, several Reynolds numbers can be defined using different denominators. This multiplicity result from the extra degrees of freedom of quantum fluids compared to classical fluid. To assess the ``instability'' of the superfluid, a possible denominator is the quantum of circulation $\kappa = h/m \simeq 0.997 . 10^{-7}\,m^2/s$ ($h$ is Planck constant and $m$ is the mass of the 4-He atom) which is the only available quantum parameter \footnote{In the literature, another denominator is sometimes used : the kinematic viscosity based on normal fluid viscosity and He-II total density ($0.90 . 10^{-8}\,m^2/s$ at $1.6\,K$\cite{DonnellyBarenghi1998}). In our case, it would result in Reynolds numbers a decade higher}. The corresponding Reynolds number $Re_{\kappa}=\Phi . V /  \kappa$ falls in the $6.10^4 - 3.10^5$ range.
\par

Fluid temperature is regulated near $1.6\,K$, which -in the two-fluid model- corresponds to $84\%$ of superfluid\cite{DonnellyBarenghi1998}. Regulation heaters are spatially distributed and located between the pump and the main honeycomb, near two doped germanium thermometers dedicated to regulation and monitoring. Cooling is provided by a pumped helium bath located above the flow. 
The excellent effective thermal conductivity of He-II makes possible an efficient regulation. The helium bath's weight pressurizes the flow and makes possible cavitation free operation.\par

\section{The second sound probe}

\begin{figure}
\onefigure[width=8cm]{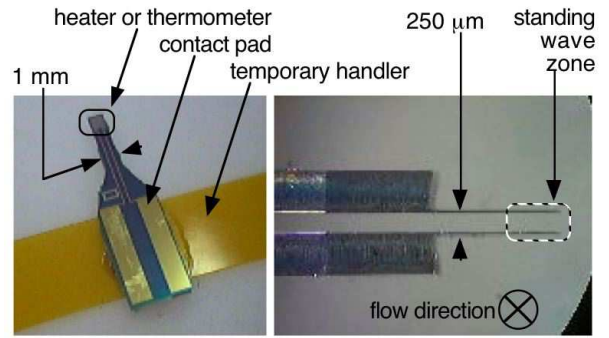} 
\caption{
Left figure : Single silicon element before the probe assembling, with contact leads and heater layer (emitter) or thermometer layer (receiver). These layers are deposited over a $0.8\,mm\,$x$\,0.8\,mm$ surface. Right figure : Side view of the tip of the assembled probe. The heater and thermometer are facing each other with a $250\,\mu m$-gap between their $15\,\mu m$-thick and $1.5\,mm$ long support. The helium mainstream flows perpendicularly to the right-side figure.
\label{fig:photo}}
\end{figure}

The second-sound miniature probe is located $80\,mm$ downstream the pipe inlet. It consists in an open cavity through which a small fraction of helium flows (see figure~\ref{fig:photo}). Both ``mirors'' of the cavity are $15\,microns$ thick silicon beams separated by a $D\simeq\,250\,\mu m$ gap.  The length and width of both beams is $1.5\,mm$ x $1\,mm$ and they are facing each other with a lateral positioning within a few tenths of mm typically. A granular Al film is deposited over a $h\simeq0.8\,mm$ square area at the tip of one beam. This film is used as transition edge superconducting thermometer. Its critical temperature is roughly $1.6\,K$. On the center of the transition, its thermal sensitivity is $T/R .\partial R / \partial T \sim 30$  and its typical resistance $350\,\Omega$. A $250\,\Omega$ parasitic resistance (film residual resistance and CuNi wiring resistance) also contributes to the total electrical resistance. Given such conditions ($250<350$), we chose to voltage bias the thermometer to prevent instabilities which can occur with a current bias across positive temperature coefficient thermistance. The typical voltage bias is a few mV. Facing this thermometer, a chromium heating film is deposited on the tip of the other beam. The cavity's volume is therefore geometrically estimated to be $\Omega \simeq h.h.D\simeq1\,mm$ x $1\,mm$ x $250\,\mu m$. Mechanical assembly and electrical connections are provided at the other ends of both beams by a stack of 3 pieces of silicon wafers. The intermediate one sets the gap of the cavity and support on both sides the contact pads. The whole geometry is designed to reduce the influence on the flow. The micromachining process, assembly and packaging of the probe will be described in details elsewhere.\par

\begin{figure}
\onefigure[width=8.5cm]{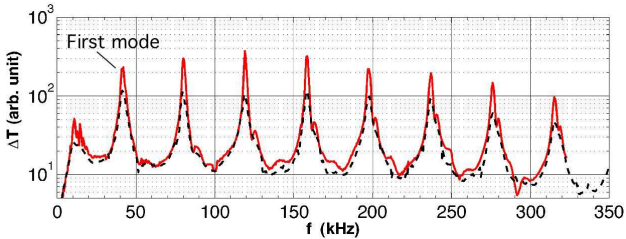} 
\caption{
The first modes of second-sound resonances of the probe, with flow (dash line) and without (solid line).
\label{fig:resonances}}
\end{figure}

Figure~\ref{fig:resonances} shows the first thermal resonances of a typical probe with the fluid at rest. Abscissa is the frequency of the applied heating and ordinate is the magnitude of the measured temperature oscillation (arbitrary units). The $n^{th}$ mode resonance frequency is in good agreement with $f_{n}=n.c/{2D}$, where $c\simeq20\,m/s$ is the second sound velocity at $1.6\,K$\cite{DonnellyBarenghi1998}. For all the probes tested, the quality factor $Q$ of the first resonance is of order $20$. It is convenient to define $\xi_{0}=\pi . f_{1} / c.Q$  as the equivalent bulk dissipation by length unit which accounts for the total attenuation on the first resonance.

Figure~\ref{fig:resonances} also shows the resonances with an He flow. Compared to the rest situation, the flow causes an extra attenuation on the temperature signal. A small fraction of it simply results from advection of the thermal wave out of the cavity. A simple ballistic model of this effect gives the equivalent bulk dissipation $\xi_{adv}=V/2h.c$ which turns out to be one order of magnitude below the signal of interest and will be neglected. The remaining of the extra attenuation is attributed to a bulk dissipation by vortex lines $\xi_{VLD}$ : this is the signal of interest. Theory predicts that a second-order resonance frequency shift is generated by vortex dissipation (\cite{Mathieu1982} and references within) but, in our case, it is much smaller than the resonance bandwidth
and we can neglect this effect. To probe the fluctuations of the VLD during experiments, the sensor is operated at the first resonance frequency. The measured temperature oscillation, called $\Delta T_{0}$ (no flow) and $\Delta T$ (with flow) are demodulated by a lock-in amplifier with a $160\,\mu s$ time constant and the oscillation's amplitude is stored for post-processing. Basic oscillator theory relates VLD dissipation $\xi _{VLD}$ and the time dependent amplitude $\Delta T$ according to\,:

\begin{equation}
\frac{\Delta T}{\Delta T_{0}}=\frac{\sinh[ \xi_{0} . D ] }{ \sinh [ (\xi_{0} + \xi _{VLD} ) . D ] }
\label{DTsurDT0}
\end{equation}

Attenuation $\xi_{VLD}$  is related to the projected vortex line density $L_{\bot}$ by\,: 

\begin{equation}
\xi_{VLD}=\frac{B.\kappa . L_{\bot}}{ 4c}
\label{xiVsLz}
\end{equation}

where the projected vortex line density $L_{\bot}$ is defined as\,:

\begin{equation}
L_{\bot}=\frac{1}{\Omega}\int {\sin^2 \theta . dl }
\label{Lz}
\end{equation}

in which $\Omega$ is the cavity volume, the summation is performed along all the vortex line elements $dl$ located inside the cavity and $\theta$ is the angle between the line elements and the axis of propagation of the second sound waves. 

Heaters in He-II are well known to induce counterflows and -for large enough heating- the counterflow can undergo self-induced turbulent transitions\cite{Tough1982, DonnellyLivreVortices,Nemirovskii1995,Melotte1998}. Consequently, a special attention has been dedicated to properly choose the heating supplied to the probe, in order to keep the probe non-invasive. With the fluid at rest, a transition has been evidenced on the temperature oscillation for a $30\,\mu W$ heating in the Cr film. The corresponding heat density, once amplified by the resonance gain $Q$  is typical of the turbulence thresholds of counterflows turbulence\cite{Tough1982, DonnellyLivreVortices,Nemirovskii1995}. During turbulence experiments, we operated the probe with driving power ranging up to $10$ times higher that the above threshold, and down to $5$ times below it. The normalized measured spectrum turns out to be independent of the driving heat, indicating that any self-induced turbulence transition in this range has a negligeable contribution on the signal. It is worth noting that Holmes and Van Sciver \cite{Holmes1992} conducted some second sound attenuation measurements with a probe heating density more than 4 decades above ours, across a He-II flow at similar temperature and velocites. These authors find average VLD fully consistent with ours.\par

To estimate the signal bandwidth, one needs to consider three time constants. First, the geometrically-limited time resolution of the probe : the time of flight  $h/V$ of order $1\,ms$. Second, the time constant of the resonator is $Q/f_{1} \simeq 0.5\,ms$, where $Q$ is the quality factor. Third, the $0.6\,ms$ time constant of the electronic set-up ($-3\,dB$ cut-off for cable/demodulation/acquisition) which has been measured without probes, using two frequency mixers
to mimic the Joule effect frequency doubling and the modulation of the signal by vortex-line-attenuation. These three time scales are comparable and the expected bandwidth is therefore $DC-1\,kHz$.\par

\begin{figure}
\onefigure[width=8.5cm]{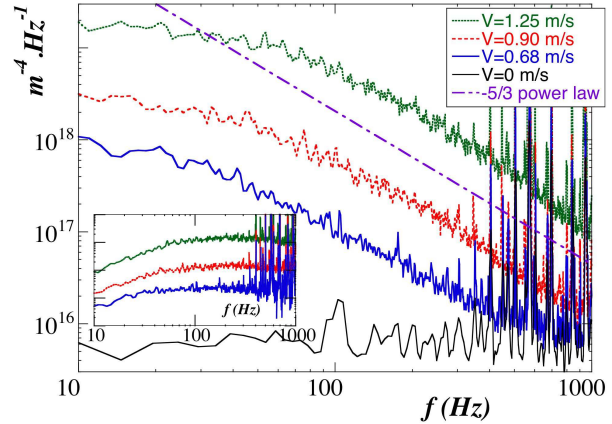} 
\caption{
Power spectrum density of the vortex line density $L_{\bot}$ for different mean flow velocities : from bottom to top $0$, $0.68$, $0.90$ and $1.25\,m/s$. The straight line is a (-5/3) power law. The insert is a $f^{-5/3}$ compensated spectrum for the 3 different mean flows after removal of a $5.10^{15}m^{-4}Hz^{-1}$ white noise floor.
\label{fig:spectre}}
\end{figure}

\section{Result}

Figure~\ref{fig:spectre} presents the main result of this letter : the power spectrum\footnote{The power spectrum density $ L_{\bot}(f).\overline{L_{\bot}(f) } $ is normalized such that its integral over positive frequencies equals $<( L_{\bot} - <L_{\bot}>)^2>$, where $<.>$ denotes time averaging. A Welch algorithm is used for spectral averaging.} of the vortex line density $L_{\bot}$ for mean flow velocities of $0$, $0.68$, $0.90$ and $1.25\,m/s$. The straight line eyes-guides the $-5/3$ power law. The insert in figure~\ref{fig:spectre} shows the 3 same spectra after subtraction of the noise floor, which is fitted from the zero-velocity spectrum by a $5.10^{15}m^{-4}Hz^{-1}$ white noise. The low frequency saturation is consistent with the integral scale plateau of order $1\,cm$. In between, over almost one decade, a local power exponent of $1.55\,\pm0.15$ is found. During other runs, this scaling has been observed down to the lowest stable mean velocity reachable in our set-up ($0.3\,m/s$). The observed dependance is compatible with a -5/3 scaling.

The inverse root of the mean VLD is often seen as a typical inter-vortex spacing. This assumes a relative smoothness of the vortex lines at small scales (negligible ``excess-length'' at scales smaller than inter-vortex spacing). With this assumption and assuming isotropy of the tangle, we find an inter-vortex spacing of $4\,\mu m$ for a mean velocity of $1\,m/s$.

We carried several tests to check the nature of the measured signal. Among these, we checked that the probe's output signal near $f_0$ wasn't modulated by residual temperature fluctuations through an hypothetical non-linear response of its thermometer.  By comparison with the signal delivered by a specially designed miniature velocity probe\cite{HaruyamaAdv1998}, we also checked that the second sound probe wasn't simply measuring velocity, for exemple through the vorticity generated in the thin boundary layers. We also checked that the measured signal wasn't history dependent, as it sometimes happens with some superfluid sensors due to trapped vortex lines\cite{Awschalom1984b}.

\section{Interpretation and Conclusion}

The interpretation of the observed spectrum is still open. As a conclusion, we discuss two approaches of it. Due to a lack of theoretical prediction and numerical observation of the projected VLD $L_{\bot}$ power spectrum, we tried in both case to relate $L_{\bot}$ to a classical counterpart. 

A first approach would amount to relate the VLD to the kinetic energy, or more precisely $L_{\bot}$ to second order statistics of the velocity components. Indeed, fourth order velocity statistics, such as power spectrum of the second order velocity components, are known to produce $-5/3$ power laws (\cite{VanAtta1975,Praskovsky1993,Hill_2001}). In a fully random tangle, the VLD is likely to be proportional to the kinetic energy (\cite{DonnellyLivreVortices} page 50-51), at least up to a log correction and assuming that the vortices are straight enough at scales smaller than the inter-vortex spacing. This linear relation can account for a $-5/3$ spectrum. Unfortunately, as pointed to us by J. Vinen, this simple relation probably no longer holds in a Kolmogorov-like tangle : the partial polarization of vortices should be strong enough to invalidate the random tangle hypothesis. A more detailed modeling is therefore necessary here.

Alternatively, the VLD can also be seen as the quantum analog of the enstrophy (vorticity square) or the absolute value of the vorticity in classical fluids\cite{VinenPRB2000}. Enstrophy in classical turbulence follows a $+1/3$ power law spectrum. Unfortunately we couldn't find any reports on the corresponding spectrum for the \textit{projected} enstrophy (projected-vorticity square) nor for the absolute value of the \textit{projected} vorticity. Nevertheless, with the sweeping theory in mind\cite{VanAtta1975,Praskovsky1993}, we can conjecture that the inertial-range dynamics of the polarization of the vortices is controled by energy containing eddies, which would result in the observed $-5/3$ power law. This picture has some analogies with passive scalar fluctuations, which also have a similar scaling.

\acknowledgments
The idea of an in-flow probe was proposed by P. Tabeling and macroscopic protoypes were done with the help of J. Maurer at the Ecole Normale Sup\'erieure. Present work results from a collaboration between three laboratories. The ESPCI designed the emitter and receiver parts of the probes, and made the Al thermometry. The ESIEE micro-machined the emitter and receiver. The CRTBT completed the probe design / machining and made the experiment and analysis.\\
P.-E.\,R. thanks Pr T. Haruyama and F. Gauthier for assistance with pressure probes, C. Lemonias and T. Fournier of NanoFab, E. Andr\'e for a bright idea on probe assembly, J. Vinen for his comments, B. Chabaud, B. H\'ebral, C. Baudet, Y. Gagne and more especially B. Castaing for numerous discussions.\\
We acknowledge financial support of Grenoble Institut de Physique de la Mati\`ere Condens\'ee, of the R\'egion Rh\^ones-Alpes and of the Agence National pour la Recherche (TSF project).

\end{document}